\documentclass[prd,twocolumn,groupedaddress,nofootinbib,preprintnumbers]{revtex4-2}

\usepackage{epsfig}
\usepackage{hyperref}
\usepackage{amssymb}
\usepackage{amsbsy}
\usepackage{amsmath}
\usepackage{url}

\newcommand{\rf}[1]{(\ref{#1})}
\newcommand{\be}{\begin{equation}}
\newcommand{\ee}{\end{equation}}
\newcommand{\bea}{\begin{eqnarray}}
\newcommand{\eea}{\end{eqnarray}}
\newcommand{\eq}[1]{Eq.~(\ref{#1})}
\newcommand{\non}{\nonumber \\*}

\newcommand{\bv}{\begin{verbatim}}
\newcommand{\ev}{\end{verbatim}}

\newcommand{\vp}{\varphi}

\newcommand{\e}{\,\mbox{e}}
\renewcommand{\d}{{\rm d}}
\renewcommand{\i}{{\rm i}}
\newcommand{\blambda}{\bar\lambda}

\newcommand{\bz}{{\bar z}}
\newcommand{\p}{\partial}
\newcommand{\bp}{\bar\partial}
\newcommand{\hg}{{\hat g}}

\newcommand{\eps}{\varepsilon}
\newcommand{\om}{\omega}

\newcommand{\LA}{\left\langle}
\newcommand{\RA}{\right\rangle}
\newcommand{\q}{Q}

\def\fun#1#2{\lower3.6pt\vbox{\baselineskip0pt\lineskip.9pt
\ialign{$\mathsurround=0pt#1\hfil##\hfil$\crcr#2\crcr\sim\crcr}}}

\begin{document}

\preprint{ITEP-TH--16/24}

\title{Nambu-Goto string as a higher-derivative Liouville theory}

\author
{Yuri Makeenko}
\vspace*{2mm}
\affiliation{NRC ``Kurchatov Institute''\/-- ITEP,  Moscow\\
\vspace*{1mm}
{makeenko@itep.ru} 
}

\begin{abstract}
 I propose a generalization of the Liouville action which corresponds to the Nambu-Goto
string like the usual Liouville action corresponds to the Polyakov string.
The two differ by higher-derivative terms which are negligible classically but revive quantumly.
An equivalence with the four-derivative action suggests that 
the Nambu-Goto string in four dimensions can be described by the (4,3) mini\-mal model
analogously to the critical Ising model on a dynamical lattice.
While critical indices are the same as in the usual Liouville theory, the domain of applicability becomes broader.
\end{abstract}



\maketitle

\section*{Introduction}

Strings are with us! A string is generically a one-dimensional object whose propagation in time forms a two-dimensional surface embedded in $d$-dimensional space-time
(we live in $d=4$). The origin of
modern string theory goes back to early 1970's when it was
recognized that the dual resonance models of strong interaction are described by strings.
Thus relativistic quantum strings do exist at the distances of the order of one fermi. There are vast
applications of strings and two-dimensional surfaces in physics: biological membranes,
cosmic strings, Abrikosov and Nielsen-Olesen vortices etc..
  
  The beauty of bosonic string theory is a simplicity of its action -- the area spanned by 
  a string propagation -- as proposed by Y.~Nambu with T.~Goto and also by H.B.~Nielsen
  at the border of 1960's and 1970's. 
  It looks very simple but this is an illusion. Area is a highly nonlinear 
  functional of $d$ target-space coordinates $X^\mu$ which is invariant under diffeomorphism 
  transformations of two coordinates parametrizing the string world-sheet. To quantize
  such a system the symmetry has to be constrained by fixing a gauge.
  The string quantization of 1970's has resulted in a 
  very beautiful theory enjoying  conformal symmetry which becomes 
  infinite-dimensional in two dimensions and whose 
  generators obey the Virasoro algebra. However, the canonical
 quantization was consistent only in $d=26$.
  
In early 1980's A.M.~Polyakov recognized that the reason for this was an additional degree of 
freedom -- one of the components of the metric tensor 
at the world-sheet -- which does not decouple if $d\neq 26$. Its dynamics is governed 
for the Polyakov string~\cite{Pol81} by the Liouville action 
(the field is accordingly called the Liouville field).
Quantization of the Polyakov string is more easy than of the Nambu-Goto string because
the  action is quadratic in $X^\mu$ and enjoys Weyl's invariance which makes conformal
symmetry manifest.
The equivalence of the Nambu-Goto and Polyakov
string formulations was shown in classical theory and at the one-loop order~\cite{FTs82} of the
perturbative expansion in 
the inverse string tension $2\pi\alpha'$. An exact solution of the quantum Liouville theory
was found by KPZ-DDK~\cite{KPZ,Dav88,DK89} using the methods of conformal field theory (CFT).
It allows to compute the so-called string susceptibility index $\gamma_{0}$ which determines
the large-area behavior of the number of surfaces of genus $h=0$. 
The result for closed Polyakov's string reads  
\be
\gamma_{0}=1-\frac{(d_+ +d_-)/2-d+\sqrt{(d_+ -d)(d_- -d)}}{12}
\label{mygstr0}
\ee
with $d_+\!=\!25$, $d_-\!=\!1$.  It is real for $d\leq1$ describing a vast amount of
models in Statistical Mechanics, in particular, $d=1/2$ describes the  susceptibility of
the critical Ising model on a random lattice~\cite{Kaz86}, but does not apply for $d>1$
where \rf{mygstr0} is not real. A pessimistic viewpoint (shared by some of my
colleagues in 1990's) is that $d=1$ is a barrier for the existence of bosonic string which
does not exist nonperturbatively if $1\!<\!d\!<\!25$ including $d=4$ or $d=3$. A more optimistic view supported
by the recent studies~\cite{DFG12,AK13,Hel14} of the spectrum 
 of ``effective strings'' is that the problem may
exist only for the Polyakov string rather than for the Nambu-Goto string. Anyway the
challenging problem of existence of nonperturbative strings is 
inherited from the previous Millennium
along with 
confinement and quantum gravity.

I argue in this Letter that \eq{mygstr0} may still hold for the Nambu-Goto string
 in $d=4$ with the KPZ barriers shifted to
 $d_\pm=15\pm 4\sqrt{6}$. Then $\gamma_0=-1/3$ like for the critical
 Ising model on a dynamical lattice~\cite{Kaz86} which for the Polyakov string was described
 by $d=1/2$. It works now in $d=4$ because $d_-\approx 5.2>4$, linking conformal symmetry
 of the Nambu-Goto string to the
 (4,3) unitary minimal model. The arguments are based on an equivalence
with the four-derivative Liouville action exactly 
 solved~\cite{Mak23e} previously. Both theories are conformal invariant in spite of the
 presence of mass parameters. 

\section*{Generalized conformal anomaly}

The standard representation of the Nambu-Goto string via auxiliary fields
which I learned from~\cite{Alv81} is
\bea
S_{\rm NG}&=&\int \sqrt{\det {(\p_a X\cdot \p_b X)}} \non &=&
\int \Big [\sqrt{g} +\frac 12\lambda^{ab} (\p_a X\cdot \p_b X-g_{ab}) \Big],
\label{SNG}
\eea
where the (imaginary) Lagrange multiplier  $\lambda^{ab}$ is a tensor density
 and  $g_{ab}$ is an independent metric tensor. I use the units where the bare string tension
 is set to 1.
 
 The action~\rf{SNG} becomes
 quadratic in $X^\mu$ that makes it easy to path-integrate $X^\mu$  out. 
For a closed string this results in an effective action
\be
{S}[g_{ab}, \lambda^{ab}]= \int \sqrt{g}
-\frac12\int  \lambda^{ab}g_{ab}
+{S}_X[g_{ab}, \lambda^{ab}],
\label{Seminva}
\ee 
where
\bea
{{\cal S}_X[g_{ab}, \lambda^{ab}] }&= &\frac d{96\pi} \int \Big[ -
\frac {12\sqrt{g}}{\tau\sqrt{\det{\lambda^{ab}}}}
+\sqrt{g}\,R\frac1 \Delta R \non &&-\Big(
\beta\lambda^{ab} g_{ab} R+2
\lambda^{ab}\nabla_a\p_b \frac1\Delta R  \Big)
\Big] 
\label{SXinv}
\eea
and terms of  higher orders in Schwinger's proper-time ultraviolet cutoff $\tau$
are dropped like they are dropped in the derivation of the Liouville action from the Polyakov string.
As argued in \cite{Mak23c} the higher-derivative terms do not change the results  in that case.
Equation~\rf{SXinv} has been derived \cite{Mak23c} from the DeWitt-Seeley expansion
of the operator $(\sqrt{g})^{-1} \p_a \lambda^{ab} \p_b$ which becomes the Laplacian for
\be
\lambda^{ab}=\blambda\sqrt{g} g^{ab}
\label{bla}
\ee 
with constant $\blambda$, reproducing the  usual Liouville action 
derived for the Polyakov string. 
Thus \eq{SXinv} generalizes the usual conformal anomaly. 
One has $\beta=1$ for the Nambu-Goto string but I keep $\beta$ arbitrary for generality.

The action \rf{SXinv} is nonlocal just as in the case of the Polyakov string. It becomes local in
the conformal gauge
\be
g_{ab}=\hg_{ab}\e^\vp,
\label{confog}
\ee
where  $\hat g_{ab}$ is the background (or fiducial) metric tensor and 
 the Liouville field $\vp$ is a dynamical variable. 
 Fixing the gauge produces the usual ghosts and their usual contribution
 to the effective action after path-integrating over the ghosts.
 A subtlety which
 will be crucial in what follows is that the curvature acquires the shift
 \be
\sqrt g R= \sqrt{\hat g} \left( \hat R - \hat \Delta \vp \right),
\label{Rshift}
\ee
where $\hat\Delta$ is the Laplacian for the metric tensor $\hg_{ab}$. It
vanishes only if the background curvature $\hat R$ vanishes.
This produces an additional nonminimal interaction of $\vp$ with background gravity. 
 
As always in Euclidean
CFT we use conformal coordinates $z$ and $\bz$ in a flat background when
$g_{zz}=g_{\bz\bz}=0$, $g_{z\bz}=g_{\bz z}=1/2$. 
Then the action \rf{Seminva} plus the ghost contribution takes the form
\bea
\lefteqn{{\cal S}[\vp, \lambda^{ab}]
=\int \e^\vp(1-\lambda^{z\bz})} \non
&&+\frac {1}{24\pi} \int \Big[ -\frac {3\e^\vp}{\tau} 
\Big(\frac{d}{\sqrt{\det{\lambda^{ab}}}}-2
\Big)
+(d-26)\vp\p\bp\vp \non &&
+d\kappa\big(2(1+\beta)\lambda^{z\bz} \p\bp \vp  
+\lambda^{zz}\nabla\p \vp +\lambda^{\bz\bz}\bar\nabla\bp\vp \big)
\Big] ,
\label{SXa}
\eea
where 
$\kappa=1$ as follows from~\rf{SXinv},  but it may be renormalized.
In the action~\rf{SXa}
$\nabla = \p-\p \vp$ is the covariant derivative in the conformal gauge and it
 describes a theory with interaction.
The representation of the $R^2$ case by an auxiliary field~\cite{KN93} is
reproduced as $\beta\to \infty$.

It is tempting to path integrate over $\lambda^{ab}$ expanding about the 
value $\blambda^{ab}=\blambda\delta^{ab}$  minimizing~\rf{SXa}
with constant $\blambda<\blambda_{\rm cl}=1$ in the world sheet coordinates
(which is determined by vanishing of $\mu^2$ in \eq{epsG} below).
However, nothing is expected to depend on $\blambda$ because of the background 
independence. 
I often keep the same notation $\lambda^{ab}$ for the fluctuations  $\delta\lambda^{ab}$
about $\blambda^{ab}$ when no confusion.
The path integral over $ \lambda^{ab}$ has a saddle point at
\be
\delta\lambda^{ab} =\sqrt{g}\tau \left(g^{ac}g^{bd}\nabla_c\p_d \vp+\frac{(\beta-1)}4 g^{ab}
 \Delta\vp\right) \frac{\kappa}{3} +{\cal O}(\tau^2)
\label{llzz}
\ee
 justified by the smallness of $\tau$. Naively
 $ \delta\lambda^{z\bz}$ is not $\sim \tau$ from
\rf{SXa}, but we should not forget  the linear in   $\delta \lambda^{z\bz}$ term
entering also the classical part of the action~\rf{SXa}, which causes the renormalization 
of the bare string tension in the scaling regime~\cite{AM16a}. 

Thus in the saddle-point approximation we arrive at the 
four-derivative action
\bea
{\cal S}[\vp]&=&\frac {1}{16 \pi b_0^2} \int \sqrt{\hg} \Big[\hg^{ab} \p_a \vp \p_b \vp
+2\mu^2 \e^\vp \non
&&+\eps  \e^{-\vp}\hat\Delta \vp \left(\hat\Delta\vp- G \hg^{ab}\,\partial_a \vp\partial_b \vp \right)\Big]  ,~~~~~
\label{inva}
\eea
where 
\begin{eqnarray}
&& b^2_0=\frac 6{26-d},\quad \mu^2=-b_0^2 \Big[8\pi \big( \bar\lambda-1 \big)+
\frac1 {\tau} \Big( \frac d{\bar\lambda}-2 \Big) \Big],\nonumber \\ && 
G=-\frac1{1+(1+\beta)^2/2},\quad
\varepsilon=-\frac {d b_0^2\kappa^2 \bar\lambda^3} {9G}\tau.
~~~~~~
\label{epsG}
\end{eqnarray}
 It  is precisely the action exactly solved in~\cite{Mak23e}.

It is clear from~\eq{llzz} that the presence of the dimensionful parameter
$\tau$ was crucial in the passage from \rf{SXa} to \rf{inva} where it becomes $\eps$. 
I refer to each of these actions as ``massive'' CFT because its  energy-momentum tensor
(EMT) will be concerved and traceless in spite of the presence of the mass parameters.

Equation~\rf{llzz} for the saddle-point values of $\lambda^{ab}$ is not the end of
the story because of the next orders in $\tau$ coming from the expansion of $1/\sqrt{\det{\lambda^{ab}}}$. But these terms are at least quartic in $\vp$
and thus are expected not to change the one-loop results while 
they may contribute to the next orders.
I shall now apply a more sophisticated technique of CFT to go toward proving the equivalence of
the actions~\rf{SXinv} and \rf{inva}.

\section*{Improved energy-momentum tensor}

 The central role in  CFT is played by the traceless EMT.
 It is derived by applying the variational derivative $\delta/\delta \hg_{ab}$ to the action in curved background which produces  terms additional to the part associated to minimal
 interaction with gravity. It was called~\cite{CCR70} ``improved'' to be distinguished from
 the minimal one and used by KPZ-DDK in solving the Liouville theory. A specifics of the
``improvement'' in two dimensions is outlined in~\cite{DJ95}.

A very nice feature of the improved EMT is that it is always traceless 
 thanks to 
the classical equation of motion for $\vp$. This is 
a general property because in the conformal gauge \rf{confog}
we have
\be
T^a_a\equiv\hat g^{ab} \frac{\delta {\cal S}}{\delta \hat g^{ab} }= -
\frac{\delta {\cal S}}{\delta \vp },
\label{tra}
\ee
where the left-hand side represents the trace of the improved EMT
while the right-hand side represents the classical equation of motion for $\vp$.

Representing $\lambda^{ab}=\sqrt{g} g^{ac}\alpha^{b}_c$ 
 with the mixed tensor 
$\alpha^b_c$ being dimensionless as required by conformal invariance and varying with respect to $\hg^{ab}$, we find  the symmetric minimal EMT
\begin{subequations}
\bea
\lefteqn{T_{zz}^{({\rm min})}=
\frac{(d-26)}{24}  (\p\vp)^2
+\frac{d\kappa}{24}\Big[2(1+\beta)\p \lambda^{z\bz}\p\vp}
\non &&+ \bp\lambda^{\bz\bz}\p\vp-\p\lambda^{\bz\bz}\bp\vp
-2 \lambda^{\bz\bz}\p\bp \vp +2\lambda^{\bz\bz}\p\vp\bp \vp 
\Big],~~~~~
 \label{Tzzmins} \\ \lefteqn{
T_{z\bz}^{({\rm min})}=
4\pi \e^\vp(1- \lambda^{z\bz})
-\frac{d\e^\vp}{2\tau\sqrt{\det{\lambda^{ab}}}}+\frac1\tau \e^\vp}\non &&
 +\frac{d\kappa}{24}\Big[\bp \lambda^{\bz\bz}\bp\vp+
 \lambda^{\bz\bz} \bp^2\vp
+\p \lambda^{zz}\p\vp + \lambda^{zz} \p^2\vp\Big].~~~~~
\label{Tzbzmins}
\eea
\end{subequations}
It is conserved obeying $\bp T_{zz}^{({\rm min})} + \p T_{\bz z}^{({\rm min})} =0$
but not traceless. 

The improved EMT is  the sum 
$T_{ab}=T_{ab}^{({\rm min})}+T_{ab}^{({\rm add})}$
of the minimal EMT and an addition, coming from the nonminimal interaction, with the component
\begin{subequations}
\bea
T_{zz}^{({\rm add})}&=& -\frac{(d-26)}{12} \p^2\vp 
-\frac{d\kappa}{24}\Big[2(1+\beta)\p^2\lambda^{z\bz} +\p\bp\lambda^{\bz\bz} \non &&
+\p \big(\lambda^{\bz\bz}\bp\vp\big)\Big]+T_{zz}^{({\rm NL})} ,\label{TzzaddL}
\\
T_{zz}^{({\rm NL})}&=&-\frac{d\kappa}{24}\Big[
\frac1{\bp} \left(\p^3 \lambda^{zz}+\p^2 (\lambda^{zz}\p\vp) \right)\Big],
~~~~
\label{TzzaddNL}
\eea
\label{Tzzadd}
\end{subequations}
where explicitly
\be
\frac 1\bp f(\om)=\int  {\d^2 z } \frac{f(z)} {\pi(z-\om)}.
\ee

In covariant notations the improved EMT reads 
\bea
T_{zz}&=&\frac1{4 b_0^2}\Big[(\p\vp)^2+2 \nabla \p \vp -2(1+\beta ) \nabla^2 \lambda^{z\bz}
-\nabla\bar\nabla \lambda^{\bz\bz}\non &&-\nabla\lambda^{\bz\bz} \bp\vp 
-2 \lambda^{\bz\bz}\p\bp\vp-4 \nabla^2 \frac 1\Delta  \nabla^2 \lambda^{zz}\Big],
\label{Tzzcov}
\eea
where we set $\kappa=6/db_0^2$ 
to simplify the formulas.
It obeys $\bp T_{z z}=0$, $T_{z \bz}=0$
thanks to the classical equations of motion. The improved
EMT is thus conserved and traceless as expected in spite of the massive parameter $\tau$! 
A price for that is the nonlocal  term \rf{TzzaddNL}. 
This is just as was discovered in~\cite{Mak22} for the action~\rf{inva}.

The conservation of the improved
EMT~\rf{Tzzcov} at the classical level follows from
\bea
\frac1\pi\bp T_{zz}&= & \p \vp \frac{\delta{\cal S}}{\delta \vp}
- \p\frac{\delta{\cal S}}{\delta \vp} -\lambda^{\bz\bz}\p 
\frac{\delta{\cal S}}{\delta \lambda^{\bz\bz}} 
+\p\lambda^{z\bz}  \frac{\delta{\cal S}}{\delta \lambda^{z\bz}}  \non &&
+ \p \big( \lambda^{zz}\frac{\delta{\cal S}}{\delta \lambda^{zz}}\big) 
+\p\lambda^{zz}\frac{\delta{\cal S}}{\delta \lambda^{zz}} .
\label{bdTzz}
\eea
In quantum theory the variations of ${\cal S}$  are no longer zeros but
 are substituted by
the variational derivatives with respect to the corresponding fields in the path integral.
For the generator of the (infinitesimal) conformal transformation $\delta z = \xi(z)$ this 
yields%
\footnote{Note  that
$\delta_\xi \lambda^{ab}=- (\p_c \xi^a) \lambda^{bc} 
- (\p_c \xi^b) \lambda^{ac}+(\p_c  \xi^c) \lambda^{ab} + \xi^c \p_c \lambda^{ab}$ 
under diffeomorphism transformations.}
\bea
\lefteqn{\hat \delta_\xi = \frac1\pi
\int \xi \bp T_{zz} = \int \Big[ \big(\xi' + \xi \p\vp\big) \frac\delta{\delta \vp}
 + \xi \p\lambda^{z\bz}  \frac\delta{\delta\lambda^{z\bz} }}\non &&
 +
\big(\xi'\lambda^{\bz\bz}  + \xi \p\lambda^{\bz\bz} \big) \frac\delta{\delta\lambda^{\bz\bz} }
 +\big(-\xi'\lambda^{zz}  + \xi \p\lambda^{zz} \big) \frac\delta{\delta\lambda^{zz} }\Big].
 \non &&
\label{hatdel}
\eea
Classically it produces
the right transformation laws of $\vp$ and $\lambda^{ab}$ whose
components $\lambda^{\bz\bz}$, $\lambda^{z\bz}$, $\lambda^{zz}$ have
conformal weights $1$, $0$, $-1$, respectively.

While the improved
EMT~\rf{Tzzcov} classically generates 
the usual conformal transformation, a short calculation gives
\be
\delta_\xi T_{zz}=\xi''' \frac1{2b_0^2}+2 \xi' T_{zz} +\xi \p T_{zz}
- \frac{1}{4b_0^2}\xi'' \frac 1{\bp} \p\nabla \lambda^{zz}
\label{nonprim}
\ee
for the transformation of the imptoved EMT itself, where the last nonlocal term is additional to the
standard transformation~\cite{BPZ} of the conserved primary%
\footnote{More accurately, it is a descendant of the unit operator.}
 tensor of the conformal
weight 2. Nevertheless I do not see any contradiction with BPZ because of the
nonlocality. Equation~\rf{nonprim} may simply show that the nonlocal term~\rf{TzzaddNL}
makes the improved EMT to be nonprimary.

The emergence of the additional nonlocal term in~\rf{nonprim} does not mean apriori 
a violation of the Virasoro algebra.
The commutator of two generators~\rf{hatdel} reads
\bea
\lefteqn{\LA\big(\hat \delta_\eta \hat \delta_\xi -\hat \delta_\xi \hat \delta_\eta\big) X \RA=
\LA \hat \delta_\zeta  X \RA} \non &&+ \frac{1}{24} \oint _{C _1}\frac{\d z}{2\pi \i} \big[ \xi'''(z)\eta(z)-\xi(z)\eta'''(z) \big]
\Big\langle \hat c X \Big\rangle
\label{Viraso}
\eea
where 
$\zeta= \xi\eta'-\xi'\eta$ in the first term on the right-hand side is as it should. 
The contour $C_1$  encircles the singularities of 
$\xi(z)$ and $\eta(z)$, leaving outside possible singularities of $X$.
As usual, $\hat c$ in \eq{Viraso} is linked to the 
 central extension of the Virasoro algebra. It is determined by the second variation of the
 action with respect to the fields as it emerges from the commutator on the
 left-hand side 
 like in~\cite{Mak22c}. It is the usual c-number for the quadratic action or at 
 one loop, but it
will be field-dependent for the action \rf{SXinv} involving the cubic part. It is
my argument in favor of the intelligent one-loop approximation described in the next Section
because $\hat c$ in \eq{Viraso} would be operator-valued otherwise.

\section*{``Massive'' versus massless CFT}

The one-loop computation of the central charge and conformal weights can be performed
by the propagators
\begin{subequations}
\bea
\LA \vp(-p) \vp(p) \RA &= &\frac {8\pi b^2}{p^2+\eps p^4}, \\
\LA \lambda^{z\bz}(-p) \vp(p) \RA &= &
 \frac{(1+\beta)G}{2}\frac {8\pi b^2 \eps}{1+\eps p^2}, \\
\LA \lambda^{zz}(-p) \vp(p) \RA &= &
 4G\frac {8\pi b^2 \eps p_\bz^2}{p^2+\eps p^4} ,
\eea
\label{propagators}
\end{subequations}
\!\!where $\eps$ and $G$ are given by \eq{epsG} and $b^2=b_0^2+{\cal O}(b_0^4)$ 
is a renormalization of $b_0^2$.
We see that $\lambda^{ab}$ has mass squared $\eps^{-1}$ and
does {\em not}\/ propagate to large distances 
what was the 
original Polyakov's agrument~\cite{Pol87} for the equivalence of the two string formulations.
However, like shown in~\cite{Mak22} for the action~\rf{inva},
 a private life of $\lambda$'s which occurs at the distances of order of the cutoff $\eps$
is seen nevertheless at macroscopic distances as a result of doing the uncertainty 
$\eps\times\eps^{-1}$ where $\eps^{-1}$ cuts momentum-space integrals and $\eps$
is the coupling of interaction between $\vp$ and $\lambda^{ab}$. This is like an appearance
of anomalies in quantum field theory (QFT).

It is clear that the local terms involving $\lambda^{ab}$ in \rf{Tzzcov}  do not contribute to the central charge because of massiveness, except for
the nonlocal term~\rf{TzzaddNL} which {\em does}\/ contribute in a full analogy 
with the four-derivative Liouville theory~\cite{Mak22}.
Its computation drastically simplifies when
 the generator of the conformal transformation is represented by \eq{hatdel} which
 accounts  for tremendous cancellations  in the quantum case, 
 while there are subtleties associated with singular products emerging in the averages
 like in the definition of the central charge $c$,
\be
\LA\hat \delta_\xi T_{zz} (0)\RA = \frac {c}{12} \xi'''(0).
\label{ccc}
\ee
The normal ordering has to be implemented in $T_{zz}$.

With  or without a little use of Mathematica we obtain 
\be
\LA\hat\delta_\xi T^{(\rm NL)}_{zz}\RA=-
\frac{1}{2b^2}
\p^2\frac 1\bp\int  d^2z \,\xi'(z)\LA\p\lambda^{zz}(z) \vp(0)\RA \delta^{(2)}(z),
\label{dTaddNL}
\ee
where
the singular product does not vanish as naively expected, 
but equals~\cite{Mak22c} 
\be
\frac1{\bp}\int \d^2 z\,  \xi'(z)\LA \p  \lambda^{zz} (z)\vp(0) \RA \delta^{(2)}(z) 
=-{G}b^2 \xi'(0).
\label{singular}
\ee
Equation~\rf{ccc}
then contributes at one loop the additional $\delta c=6G$ to the central charge 
 which is the same as for the four-derivative action~\rf{inva}.
 
It is tempting to repeat the arguments~\cite{Mak23e} that like for the action~\rf{inva} the intelligent one loop will give
an exact answer in our case as well. I call this way the procedure proposed 
by DDK~\cite{Dav88,DK89}
for solving the usual Liouville theory where $T_{zz}^{({\rm add})}$ is simply multiplied by
a parameter $\q$ describing a renormalization of the nonminimal interaction.
The arguments rely on a cancellation of skeleton diagrams which is 
represented by \eq{hatdel}.
We then would obtain $\delta c=6\q G$ for the additional contribution to the central charge at all loops.
The vanishing of the total central charge would then require
\be
d-26+\frac{6\q^2}{b^2}+1+6\q G=0
\label{DDK1}
\ee
recovering DDK for $G=0$. 

The second equation that fixes the conformal weight of $\e^\vp$ 
to be 1 remains unchanged 
\be
1=\q- b^2
\label{DDK2}
\ee
what is easily seen from the propagators \rf{propagators} when $\eps\to0$.
Like for the four-derivative action
the nonlocal term in $T_{zz}$ does not contribute to the conformal weight.
In fact there exists a whole family of primary operators with the weights 1
thanks to \eq{DDK2},  including a renormalized version of $\e^\vp/\sqrt{\det \lambda^{ab}}$.

The difference between massless and ``massive'' CFTs is explicitly seen in the pure $R^2$ case
 where it can be illustrated by the computation of the average~\rf{ccc} at large $\beta$.
 Then only $\vp$  and $\lambda^{z\bz}$ are essential and we obtain
\bea
\LA\hat\delta_\xi T_{zz}\RA&=& \frac1{4b^2} \Big\{2\q^2+\!\int\! d^2 z \Big[ \xi'(z) \p^2 \vp(z)
+\xi(z) \p^3 \vp(z)  \non
&&+(1+\beta)
\big(2\xi''(z) \p\lambda^{z\bz}(z)-6\xi'(z) \p^2\lambda^{z\bz}(z) \non
&&-4\xi(z) \p^3\lambda^{z\bz}(z) \big)\Big]\vp(0) \delta^{(2)}(z)  \Big\}.
\label{dTmin}
\eea
with the singular products equal to
\begin{subequations}
\bea
&&\int \d^2 z\, \xi(z) \LA \p^n  \vp(z) \vp(0)\RA  \delta^{(2)}(z)
 =G   b^2  (-1)^{n}\p^{n}\xi(0) S_{n} , \non &&\label{singvp}\\ 
&&\int \d^2 z\,  \xi(z)\LA \p^n  \lambda^{z\bz}(z) \vp(0)\RA  \delta^{(2)}(z) 
\nonumber \\[-1mm]&&~~~~~~~~~
=\frac12{G   b^2  (1+\beta)}
 (-1)^{n}\p^{n}\xi(0)T_{n} . 
\eea
\label{singular}
\end{subequations}
In the free massless case of $\eps=\infty$ like in~\cite{KN93}
we  have  
\be
S_{n}=0,\quad T_{n}=\frac{2}{n(n+1)}\qquad \fbox{massless}
\ee
while in the massive case
\be
S_{n}=\frac{2}{n(n+1)},\quad T_{n}=\frac{2}{(n+1)}\qquad \fbox{massive}.
\ee
The derivation is as in~\cite{Mak22c}.  
Substituting into \eq{dTmin}, we see  in the massless case 
the appearance in the central charge $c$ of 2 rather than 1 as for the usual Liouville action due to
 the presence of two orthogonal scalars $\vp\!+\!(1\!+\!\beta)\lambda^{z\bz}$ and $\lambda^{z\bz}$~\cite{KN93}. 
However, in our case of  $\eps\to0$ only  the first combination remains massless  and contributes 1 to $c$, thus demonstrating the $R^2$-term does not change the
usual results.
 In ``massive'' CFT conformal symmetry holds for all distances, not only for the distances
 $\gg \sqrt{\eps}$ where the standard CFT technique of BPZ~\cite{BPZ} applies.
 In contrast to CFT without diffeomphism invariance,
 the value of $\eps$ can now be compensated by a shift of $\vp$.
 The solution to Eqs.~\rf{DDK1}, \rf{DDK2} will be  immediately  described.

\section*{Relation to minimal models}

For the Nambu-Goto string we have $G=-1/3$ from \rf{epsG}.
Keeping in mind 
applications of the four-derivative Lioville action~\rf{inva} in other cases,
let us consider arbitrary $G$.
From Eqs.~\rf{DDK1}, \rf{DDK2}  $b_0^{-2}$ renormalizes to
\begin{subequations}
\bea
b^{-2}& = &
\frac {13-d-6 G+\sqrt{(d_+-d)(d_--d)}}{12}, ~~~
\label{tb0}\\
d_\pm&=&13-6 G\pm 12 \sqrt{1+G} 
\label{dmp}
\eea
\label{tb}
\end{subequations}
and $\gamma_{0}=1-b^{-2}$  ($\gamma_h=1+(h-1)b^2$ for surfaces of genus $h$) is
 as in \eq{mygstr0} with 
the KPZ barriers shifted to $d_\pm$ given by \rf{dmp}.

The values of $d_\pm$ depend on $G$ which has to lie in the interval $[-1,0]$ for the  
 the action~\rf{inva} to be stable.
Then $b^{-2}$ is real for $d<d_-$ which increases from 1 at $G=0$ to 19 at $G=-1$.
For $G=-1/3$  we have 
$
d_-=15-4\sqrt{6}\approx 5.2 >4
$
as is already announced in the Introduction,
so $\gamma_{0}$ is real in $d=4$. Remarkably, the value
 $G=-1/3$ is associated in $d=4$ with the $p=3$, $q=p+1=4$ unitary minimal model as
 it will be momentarily discussed.

To find the relation to minimal models we note that 
the operators
\be
V_\alpha=\e^{\alpha \varphi}, \qquad \alpha=\frac{1-n}2+\frac{1-m}{2b^2}
\ee
are the BPZ null-vectors for integer $n$ and $m$ like in the usual Liouville theory~\cite{ZZ}.  
Their conformal weights 
\be
\Delta_\alpha =\alpha+(\alpha-\alpha^2) b^2
\ee
are derived like \rf{DDK2} and
reproduce Kac's spectrum of CFT with the central charge
\be 
c=1+6(b+b^{-1})^2,
\label{cb}
\ee
where $b$ is given by \eq{tb}. This $c$ is the central charge 
of the Virasoro algebra  and not to be confused with the central 
charge $c^{(\vp)}=26-d$ of $\vp$ (or $\vp$ plus  $\lambda$'s). 

The minimal models are obtained by choosing
\be
c=25+6 \frac {(p-q)^2}{pq}
 \label{c25}
\ee
with coprime $q>p\geq 2$. Then Eqs.~\rf{tb}, \rf{cb} fix
\be
G=\frac {(1-d-6 \frac {(p-q)^2}{pq})q}{6(q+p)}, ~~
1-6\frac{(p-q)^2}{pq}\leq d\leq 19-6\frac pq.
\label{cG}
\ee
In the usual Liouville theory where $G=0$ 
\eq{cG} implies $d=26-c=1-6 \frac {(p-q)^2}{pq}$ for
the central charge of matter, but   $d$ is 
a free parameter for $G\neq 0$. Contrary to the Liouville theory now  $c\neq 26-d$.
The inequalities in \rf{cG} guarantee $0\geq G \geq -1$ as is necessary for stability. 

Given \rf{cG}  we finally find  from~\eq{tb}
\be
b^{-2}=\left\{ 
\begin{array}{ll}
\displaystyle{\frac qp}&\hbox{perturbative branch}\\[4mm]
\displaystyle{-1+\frac{(25-d)p}{6(q+p)} }\quad &\hbox{the other branch}\end{array} 
\right. 
\ee
that applies for $d>25 -6{\frac {(p+q)^2}{p^2}}\geq2$.
The perturbative branch is as in the usual 
Liouville theory, but the second branch is no longer 
$p \leftrightarrow q$ interchangeable
with it. 

There are no obstacles against $d=4$ for the unitary case~\cite{Uni} $q=p+1$!
The barriers $d_\pm$ coincide (both equal 19) for
$ d=d_{\rm c}=13-\frac 6p  $ ($d_{\rm c}\geq 10$ for $p\geq2$).
For $d$ from the interval $1\leq d<d_{\rm c} $  we have $d\leq d_-$ and $\gamma_{0}$ is {\em real}.

\section*{Final remarks}

While our critical indices are as for the perturbative branch in the usual 
Liouville theory, the domain of applicability is now broader complimenting 
 applications of the Liouville action in Condensed Matter Theory inspired by~\cite{KMT96}.
 A hint that conformal symmetry of the Nambu-Goto string and the
 critical Ising model
on a dynamical lattice may both be described by the (4,3) minimal model sounds 
 fascinating and deserving future studies.
 
 Equation~\rf{SXinv} generically represents 
a ``massive'' generalization of the Liouville action. 
I expect it may have wide applications
and help to make an insight in quantum gravity as the usual Liouville action did~\cite{Rie84,FTs84}.

The reason why in \eq{SXinv} I have dropped the terms of higher order in $\tau$,
which has resulted in the four-derivative action~\rf{inva}, 
is twofold. First, this is a simplest extension of the Liouville action, 
demonstrating how the higher-derivative terms revive. Second, there could be a kind
of universality of the higher terms like it happens for the Polyakov string
where they are marginal and
 do not change~\cite{Mak23c} the results of the usual Liouville action.

I can illustrate this by showing how the nonlocal term~\rf{TzzaddNL} emerges
for the Nambu-Goto string in flat space where it precisely comes from averaging 
over $X^\mu$ of EMT for the Nambu-Goto action~\rf{SNG}
\be
T_{zz}=2\pi\LA \lambda^{z\bz}\p X\cdot \p X +\lambda^{\bz\bz}\p X\cdot \bp X  \RA
\label{Xzz}
\ee
thanks to the interaction $ \lambda^{zz}\p X\cdot \p X$ in~\rf{SNG}.
The emerging $T_{zz}$ is represented by the diagrams 
\vspace*{4mm}
\begin{figure}[h]
\includegraphics[width=8.5cm]{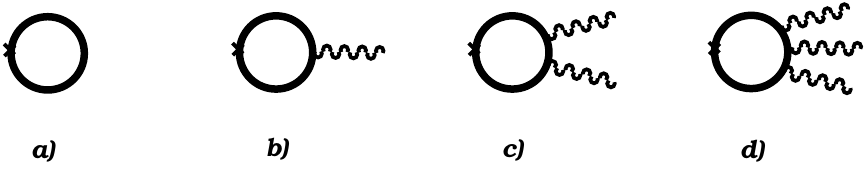} 
\caption{Diagrams representing the average of  \rf{Xzz} over $X^\mu$.
The solid line represents $X^\mu$ and the wavy lines represent $\lambda^{ab}$.}
\label{mean-F1}
\end{figure} 
depicted in Fig.~\ref{mean-F1}.
For the diagram in Fig.~\ref{mean-F1}$b)$ we find precisely 
the linear part of the nonlocal term~\rf{TzzaddNL}.
For the diagram in Fig.~\ref{mean-F1}$c)$ nonlocality emerges only 
when the two $\lambda$'s are
$\lambda^{zz}$ and $\delta\lambda^{z\bz}$. We obtain analogously to~\rf{TzzaddNL}
\bea
\hbox{Fig.~\ref{mean-F1}$c)$}=
-\frac{d \kappa}{12}\,\frac1{\bp}\p^3\big(\lambda^{zz} \delta\lambda^{z\bz} \big)
\label{NLll}
\eea
which may be thought as obtained from a six-derivative action.
Nonlocal terms emerging from the diagrams in Fig.~\ref{mean-F1}d) etc. will not 
contribute at one loop.
The expectation is that the sum of the contributions of~\rf{TzzaddNL} and \rf{NLll}
to $\langle\hat\delta_\xi T_{zz}\rangle$ with the singular products computed for
the six-derivative action may give the same result as \rf{dTaddNL} computed for
the four-derivative action, like  it was the origin of the universality 
of higher terms for the Polyakov string~\cite{Mak23c}. I hope to return to this
issue elsewhere.

\begin{acknowledgements}
I  thank J.~Ambj\o rn, A.~Barvinsky, K.~Dasgupta, V.~Kazakov, 
C.~Kristjansen, A.~Morozov, G.~Semenoff, M.~Shaposhnikov,  A.~Tseytlin, 
D.~Vasiliev, M.~Vasiliev, A.~Zamolodchikov, K.~Zarembo for 
useful comments. 
This work was carried out within the research program of NRC ``Kurchatov Institute''.
\end{acknowledgements}

\end{document}